\documentclass[conference]{IEEEtran}
\IEEEoverridecommandlockouts

\usepackage{cite}
\usepackage{amsmath,amssymb,amsfonts}
\usepackage{algpseudocode}
\usepackage{graphicx}
\usepackage{textcomp}
\usepackage{xcolor}
\usepackage{graphicx}
\usepackage{booktabs}
\usepackage{tcolorbox}
\usepackage{xspace}
\usepackage{algorithm}
\usepackage{algpseudocode}
\usepackage{subcaption}
\usepackage{tikz}
\usetikzlibrary{decorations.pathreplacing}
\usepackage{hyperref}

\newcommand{\wavycoverbandauto}[7]{%
  \path (#1.south west);
  \pgfgetlastxy{\imgwest}{\imgbottom}
  \path (#1.north east);
  \pgfgetlastxy{\imgeast}{\imgtop}

  \pgfmathsetmacro{\yb}{\imgbottom/1cm}%
  \pgfmathsetmacro{\yt}{\imgtop/1cm}%

  \pgfmathsetmacro{\xcenter}{#2}%
  \pgfmathsetmacro{\halfwidth}{#3}%
  \pgfmathsetmacro{\amp}{#4}%
  \pgfmathsetmacro{\period}{#5}%
  \pgfmathsetmacro{\overhang}{#7}%

  \pgfmathsetmacro{\ystart}{\yb-\overhang}%
  \pgfmathsetmacro{\yend}{\yt+\overhang}%

  \fill[white]
    plot[
      domain=\ystart:\yend,
      samples=180,
      variable=\y
    ]
    ({\xcenter - \halfwidth + \amp*sin(360*(\y-\ystart)/\period)}, {\y})
    --
    plot[
      domain=\yend:\ystart,
      samples=180,
      variable=\y
    ]
    ({\xcenter + \halfwidth + \amp*sin(360*(\y-\ystart)/\period)}, {\y})
    -- cycle;

  \draw[gray, line width=#6]
    plot[
      domain=\ystart:\yend,
      samples=180,
      variable=\y
    ]
    ({\xcenter - \halfwidth + \amp*sin(360*(\y-\ystart)/\period)}, {\y});

  \draw[gray, line width=#6]
    plot[
      domain=\ystart:\yend,
      samples=180,
      variable=\y
    ]
    ({\xcenter + \halfwidth + \amp*sin(360*(\y-\ystart)/\period)}, {\y});
}
\newcommand{\wavycoverbandrel}[7]{%
  \path (#1.south west);
  \pgfgetlastxy{\imgwest}{\imgbottom}
  \path (#1.north east);
  \pgfgetlastxy{\imgeast}{\imgtop}

  \pgfmathsetmacro{\xcenter}{(\imgwest + (#2)*(\imgeast-\imgwest))/1cm}%
  \wavycoverbandauto{#1}{\xcenter}{#3}{#4}{#5}{#6}{#7}%
}

\DeclareRobustCommand{\figmark}[1]{%
  \tikz[baseline=(char.base)]{%
    \node[
      shape=circle,
      draw,
      inner sep=0pt,
      minimum size=1em,
      line width=0.35pt,
      font=\scriptsize
    ] (char) {#1};%
  }%
}

\newboolean{showcomments}
\setboolean{showcomments}{true} 
\ifthenelse{\boolean{showcomments}}{
  \newcommand{\nbc}[3]{
    {\colorbox{#3}{\bfseries\sffamily\scriptsize\textcolor{white}{#1}}}%
    {\textcolor{#3}{\textsf\small$\blacktriangleright$\textit{#2}$\blacktriangleleft$}}}
  \providecommand{\todo}[1]{\nbc{TODO}{#1}{blue}\xspace}
}{
  \newcommand{\nbc}[3]{}
  \renewcommand{\todo}[1]{}
}

\newcommand{\takuto}[1]{\nbc{Takuto}{#1}{cyan}\xspace}
\newcommand{\rk}[1]{\nbc{Raula}{#1}{orange}\xspace}

\makeatletter
\let\orig@lstnumber=\thelstnumber

\newcommand\lstresetnumber{\global\let\thelstnumber=\orig@lstnumber}
\makeatother

  \definecolor{mygray}{gray}{0.6}

\def\BibTeX{{\rm B\kern-.05em{\sc i\kern-.025em b}\kern-.08em
    3T\kern-.1667em\lower.7ex\hbox{E}\kern-.125emX}}
    
\begin{document}
\title{Agent ATO: Visualizing Agent Interaction Timelines from Logs} 

 \author{
 \IEEEauthorblockN{
 Takuto Kawamoto\IEEEauthorrefmark{1},
 Yoshiki Higo\IEEEauthorrefmark{1},
 Raula Gaikovina Kula\IEEEauthorrefmark{1}
 }
 \IEEEauthorblockA{\IEEEauthorrefmark{1}The University of Osaka, Japan}
 }

\maketitle

\begin{abstract}
AI coding agents are becoming part of developers' workflows, but their behavior is difficult to understand from final code changes alone.
During a task, agents interact with software repositories through sequences of actions such as searching for files, reading code, editing programs, and running tests or build commands.
These interactions, together with token usage, are often recorded in console logs, but raw logs are difficult for developers to inspect.
In this paper, we propose Agent ATO (Agentic Trajectory Observer), a tool for visualizing AI coding agent interaction timelines from console logs.
Agent ATO extracts agent interactions, classifies them by command or tool type, and visualizes them as timelines.
In addition to an all-interaction timeline, Agent ATO provides filtered timelines that emphasize file discovery, file reading, file editing, and execution while preserving surrounding context.
We illustrate how Agent ATO may help developers inspect and compare agent actions using selected runs from two repair tasks.
Future work will apply Agent ATO to more agents, tasks, and development environments, and will evaluate whether it reduces the effort needed to compare trajectories.
\end{abstract}


\section{Introduction}
\label{1_introduction}

AI coding agents are increasingly used to automate software engineering tasks by interacting with development environments.
Unlike one-shot uses of large language models, these agents iteratively search repositories, inspect files, edit code, and run tests through an agent-computer interface~\cite{sweAgent}.
Their work processes are often recorded in console logs, which contain tool calls, commands, file inspections, edits, execution results, and errors.
These logs can reveal how the agent proceeded toward final code changes, such as whether it explored relevant files before editing, validated its changes, or returned to exploration after failures.

\begin{figure*}[t]
  \centering
  \begin{tikzpicture}
    \node[anchor=south west, inner sep=0] (img) at (0,0) {%
      \includegraphics[width=\textwidth]{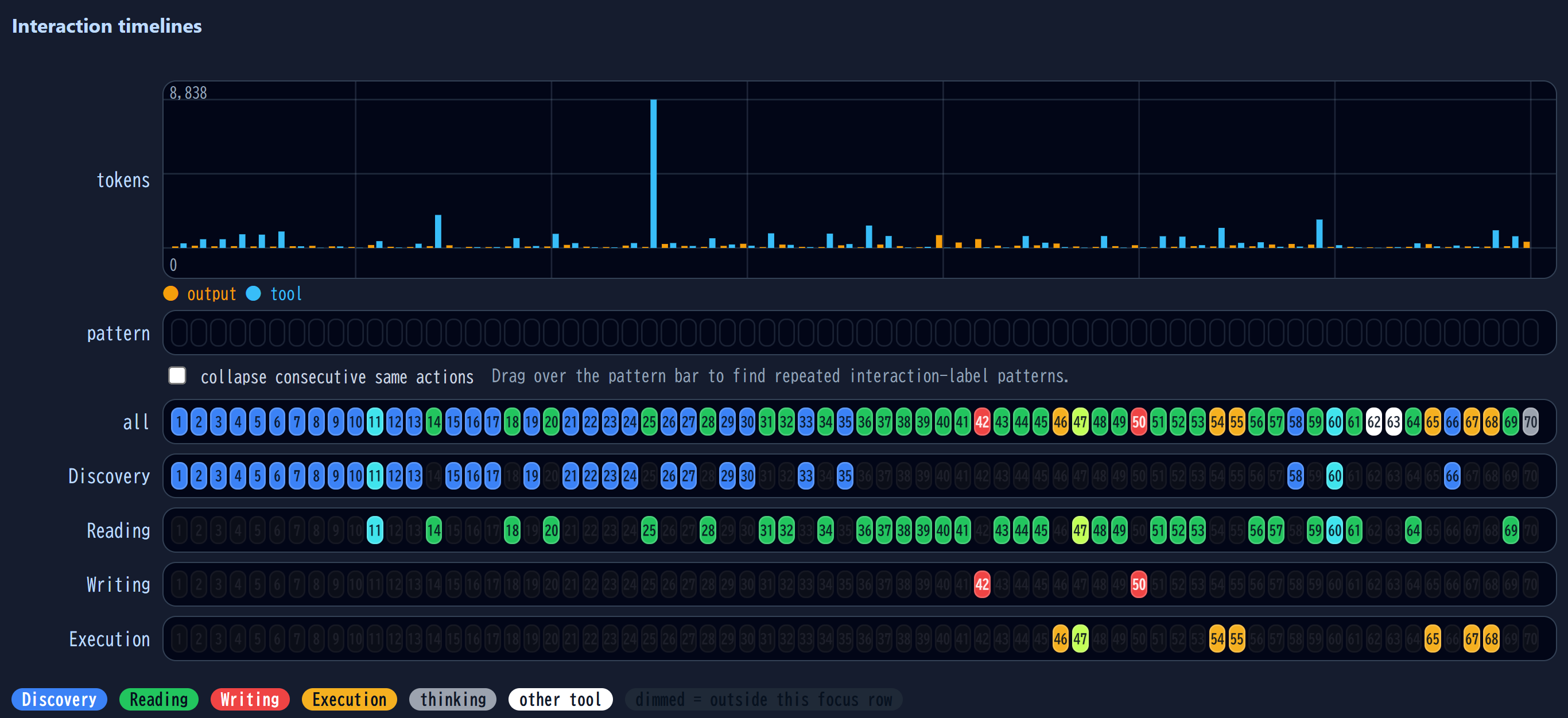}
    };

    \begin{scope}[x={(img.south east)}, y={(img.north west)}]
      \tikzset{
        anno/.style={
          circle,
          draw=yellow,
          fill=none,
          text=yellow,
          font=\bfseries\small,
          inner sep=1.2pt,
          minimum size=4.5mm
        }
      }

      \node[anno] at (0.030,0.410) {1};
      \draw[yellow, thick, ->] (0.045,0.410) -- (0.072,0.410);

      \node[anno] at (0.014,0.245) {2};
      \draw[
        yellow,
        thick,
        decorate,
        decoration={brace, amplitude=5pt, mirror}
      ] (0.038,0.35) -- (0.038,0.095);

      \node[anno] at (0.04,0.845) {3};
      \draw[yellow, thick, rounded corners=2pt]
        (0.060,0.562) rectangle (0.992,0.895);
    \end{scope}
  \end{tikzpicture}
  \caption{Overview of our prototype visualization. Each small rectangle represents an interaction reconstructed from the Pi coding agent event log.}
  \label{fig:prototype}
\end{figure*}
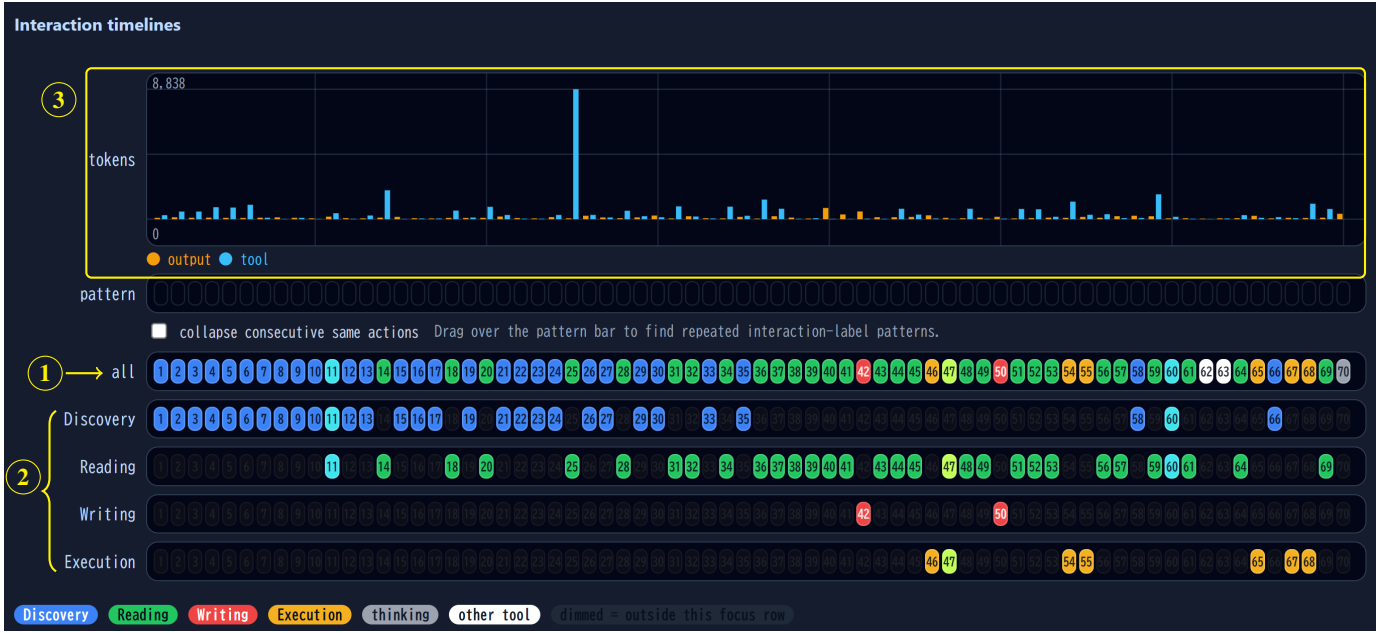

Prior studies have analyzed agent trajectories and traces to understand agent behavior. Some work examines software engineering agent trajectories through thought-action-result sequences and execution traces in automated program repair agents~\cite{understandingTrajectories,understandingThroughTraces}. Other work analyzes trajectories to identify why LLM agent systems fail~\cite{whyFail}. Tool support has also been motivated by the need to debug agent workflows and review long agent histories~\cite{agentDebugger}. In parallel, visual tools for LLMs have compared responses across prompts or trials for prompt improvement~\cite{promptIDE,promptAid,chainForge}. Recent work on coding agents also shows that comparing agent behavior across trials is a relevant visual analytics problem~\cite{illuminating}.

However, although console logs record events in temporal order, they do not clearly present interaction types or their ordering.
Developers and researchers must still read long logs to locate searches, reads, edits, executions, and segments requiring closer analysis.
This makes it difficult to notice process-level cues, such as missing validation after edits or no further exploration after failures. 
A clearer view of agent execution can help developers inspect past runs and refine future runs.
Similar problems have arisen in other software-development contexts, where event logs have been aggregated and visualized to support behavioral understanding, such as in debugging activity analysis~\cite{blueprint}.

We propose \textit{Agent ATO}, a timeline visualization that transforms console logs into classified interaction sequences for inspecting and comparing agent trajectories. We focus on interaction histories observable from console logs. In this paper, an interaction is a coherent unit of agent activity initiated by an LLM response and realized through tool use, such as repository navigation, file reading, file editing, or command execution. A sequence of interactions forms an agent trajectory. Agent ATO shows an overview timeline of all interactions and filtered timelines for four tags: Discovery, Reading, Writing, and Execution. Because a single interaction may belong to multiple tags, the filtered timelines let users focus on one perspective while preserving temporal context. We implemented a prototype for the Pi coding agent~\footnote{\url{https://github.com/earendil-works/pi}}.

To demonstrate Agent ATO, we conduct two case studies using repeated executions of the same coding task under the same prompt and environment. The first case study shows that executions with similar repair directions can differ in validation behavior: one attempt formed repeated edit-validation cycles, whereas another edited without corresponding test execution. The second case study shows that high token usage can have different causes: some spikes come from long LLM-output segments that repeatedly reconsider repair policies, while others come from large tool results produced by file-reading operations. Together, these cases show how Agent ATO helps users compare trajectories and identify process-level cues that are difficult to see from final patches or raw logs alone.

This paper makes the following three contributions.
First, we define AI coding agent behavior as an interaction trajectory composed of observable software-development operations.
Second, we propose a timeline visualization that combines detailed command/tool types with filtered views based on Discovery, Reading, Writing, and Execution.
Third, we demonstrate through two case studies how the visualization supports comparison of repeated executions, exposes differences in edit-validation behavior, and distinguishes token usage caused by LLM output from token usage caused by tool results.
Rather than automating diagnosis, the visualization provides a basis for observing and comparing AI coding agent trajectories.

\section{Agent ATO (\textbf{A}gentic \textbf{T}rajectory \textbf{O}bserver)}
\label{2_approach}

Fig.~\ref{fig:prototype} shows the Agent ATO visualization.
Each rectangle represents one interaction, ordered from left to right.
We call this sequential representation a timeline, although it does not represent elapsed time or duration.
The row marked \figmark{1} is the all-interaction timeline, which presents the complete trajectory.
The rows marked \figmark{2} are filtered timelines for Discovery, Reading, Writing, and Execution.
These timelines highlight interactions for one analytical perspective while keeping the remaining interactions visible in gray for context.
The graph marked \figmark{3} shows token usage aligned with the same interaction sequence and separates LLM-output tokens from tool-result tokens.
We describe these three parts below.

\subsection[]{\figmark{1} All-interaction timeline}
The all-interaction timeline provides the complete sequence of reconstructed interactions for a task.
Agent ATO takes event logs collected during AI coding agent executions as input.
These logs include LLM messages, tool-call events, bash commands, tool results, token usage, and timestamps.
Agent ATO uses these events to reconstruct the order of commands and tool uses while retaining detailed messages and outputs for drill-down inspection.

Each mark represents one interaction, and the marks are arranged from left to right according to execution order.
By scanning the sequence, users can see whether the agent explored and read files before writing, whether writing occurred early, and whether execution followed writing.

\subsection[]{\figmark{2} Filtered timeline }

The filtered timelines reuse the same interaction sequence but emphasize four analytical perspectives:

\begin{itemize}
\item \textbf{Discovery} highlights operations for finding relevant files or information, such as \texttt{ls}, \texttt{find}, \texttt{rg}, and \texttt{grep -l}. These interactions are shown in blue.
\item \textbf{Reading} highlights operations for inspecting file contents or diffs, such as \texttt{read}, \texttt{cat}, \texttt{grep}, and \texttt{git diff}. These interactions are shown in green.
\item \textbf{Writing} highlights operations that modify files or repository state, such as \texttt{edit}, \texttt{write}, and \texttt{rm}. These interactions are shown in red.
\item \textbf{Execution} highlights validation or execution operations, such as running tests, builds, linters, type checkers, or scripts. These interactions are shown in yellow.
\end{itemize}

Interactions that do not match the selected perspective remain visible in gray.
This design lets users focus on one view, such as Reading or Execution, without losing the surrounding temporal context.
When an interaction belongs to multiple perspectives, Agent ATO shows it using a blended color derived from the corresponding tags.

\begin{figure*}[!t]
  \centering
  \begin{tikzpicture}
    \def\W{15.0}
    \def\H{3.2}
    \node[anchor=south west, inner sep=0] (img) at (0,0) {%
      \includegraphics[
        width=15.0cm,
        trim=0pt 92pt 0pt 0pt,
        clip
        ]{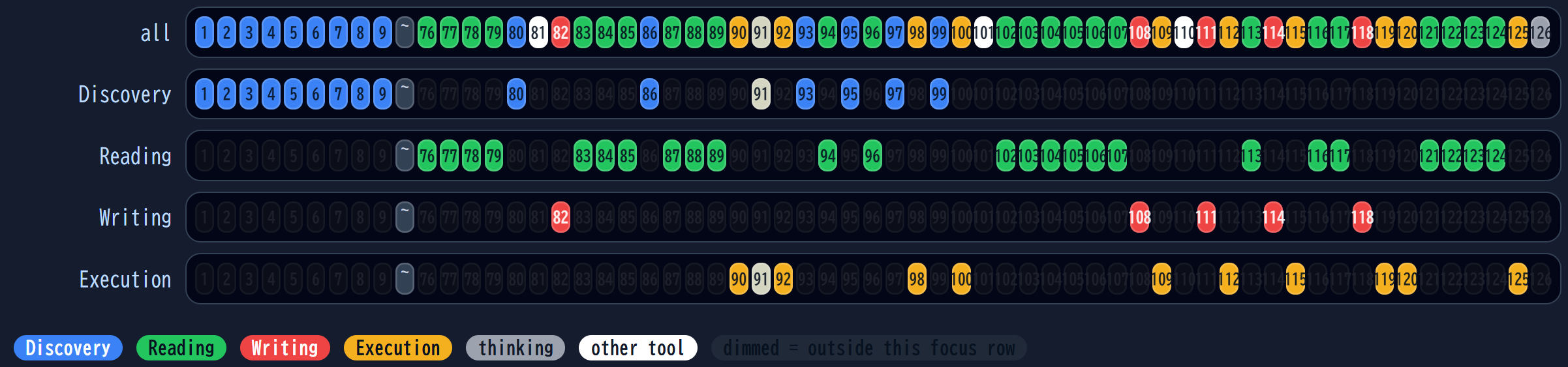}%
    };
    \wavycoverbandrel{img}{0.25}{0.32}{0.06}{1.00}{0.45pt}{0.40}
  \end{tikzpicture}
  \caption{Case~1, Attempt~1. Writing interactions are followed by Execution interactions, showing repeated edit-validation cycles.}
  \label{fig:case1-timeline1}
\end{figure*}

\begin{figure*}[!t]
  \centering
  \begin{tikzpicture}
    \def\W{15.0}
    \def\H{3.2}
    \node[anchor=south west, inner sep=0] (img) at (0,0) {%
      \includegraphics[
        width=15.0cm,
        trim=0pt 92pt 0pt 0pt,
        clip
        ]{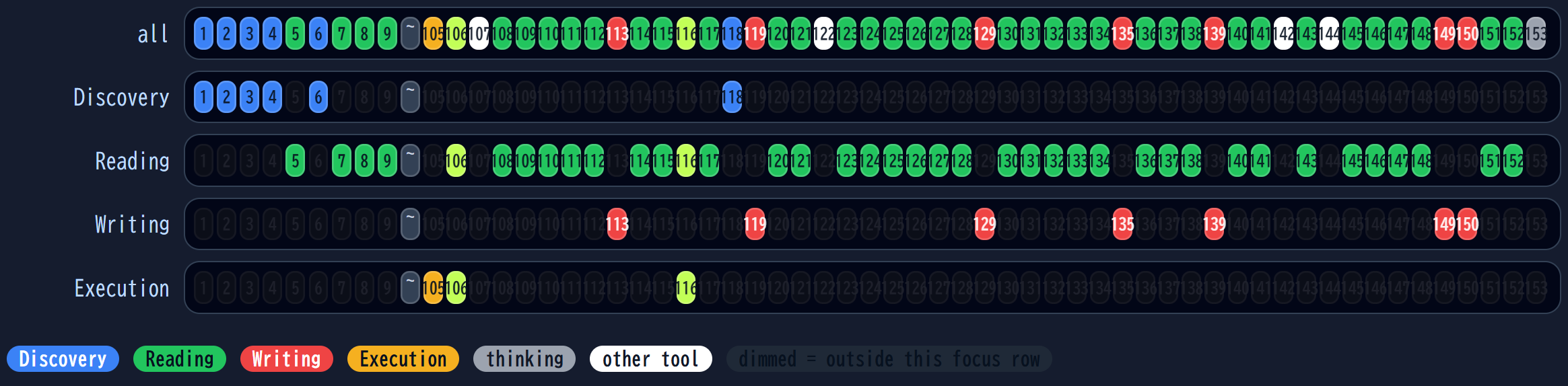}%
    };
    \wavycoverbandrel{img}{0.25}{0.32}{0.06}{1.00}{0.45pt}{0.40}
  \end{tikzpicture}
  \caption{Case~1, Attempt~2. The run contains several Writing interactions after context gathering, but these edits are not followed by corresponding Execution interactions.}
  \label{fig:case1-timeline2}
\end{figure*}

\subsection[]{\figmark{3} Token-usage View }

The token-usage graph is aligned with the same interaction sequence as the timelines.
For each interaction, it separates token usage into LLM-output tokens and tool-result tokens.
This distinction helps users tell whether high token usage came from a long agent response or from a large tool result.
Because the graph is aligned with the operation sequence, users can inspect not only where token usage is high, but also what type of activity caused it.

\section{Technical Details}
In this section, we describe the technical implementation and user interactions behind the visualization. 

\subsection{Implementation}
We implemented a prototype consisting of two components: a TypeScript extension for collecting Pi coding agent events and a Python script for generating the visualization.
The TypeScript extension records events during Pi agent execution and saves them as a JSONL file.
The recorded events include Pi turn boundaries, LLM messages, tool-call events, bash commands, tool results, token usage, and timestamps.

In Pi, a turn consists of one LLM response and the tool calls triggered by that response.
A single user prompt may therefore produce multiple Pi turns.
In our implementation, we treat each interval from \texttt{turn\_start} to \texttt{turn\_end} as one interaction.
For each interaction, the generation script assigns a representative label based on the executed bash command, pipeline, or tool name.
For example, it uses labels such as \texttt{find | grep | head}, \texttt{git diff}, or \texttt{npm test} to summarize the concrete operation performed in the interaction.

The script also assigns one or more tags---Discovery, Reading, Writing, and Execution---to each interaction based on rule-based matching of commands and tool names.
It then generates a static HTML file that contains the all-interaction timeline, filtered timelines, token-usage graph, and interactive detail views.
The generated HTML file can be opened in a web browser, allowing users to inspect trajectories without running a separate visualization server.

\begin{figure*}[!t]
  \centering
  \includegraphics[
    width=\linewidth,
    trim=0pt 630pt 0pt 0pt,
    clip
  ]{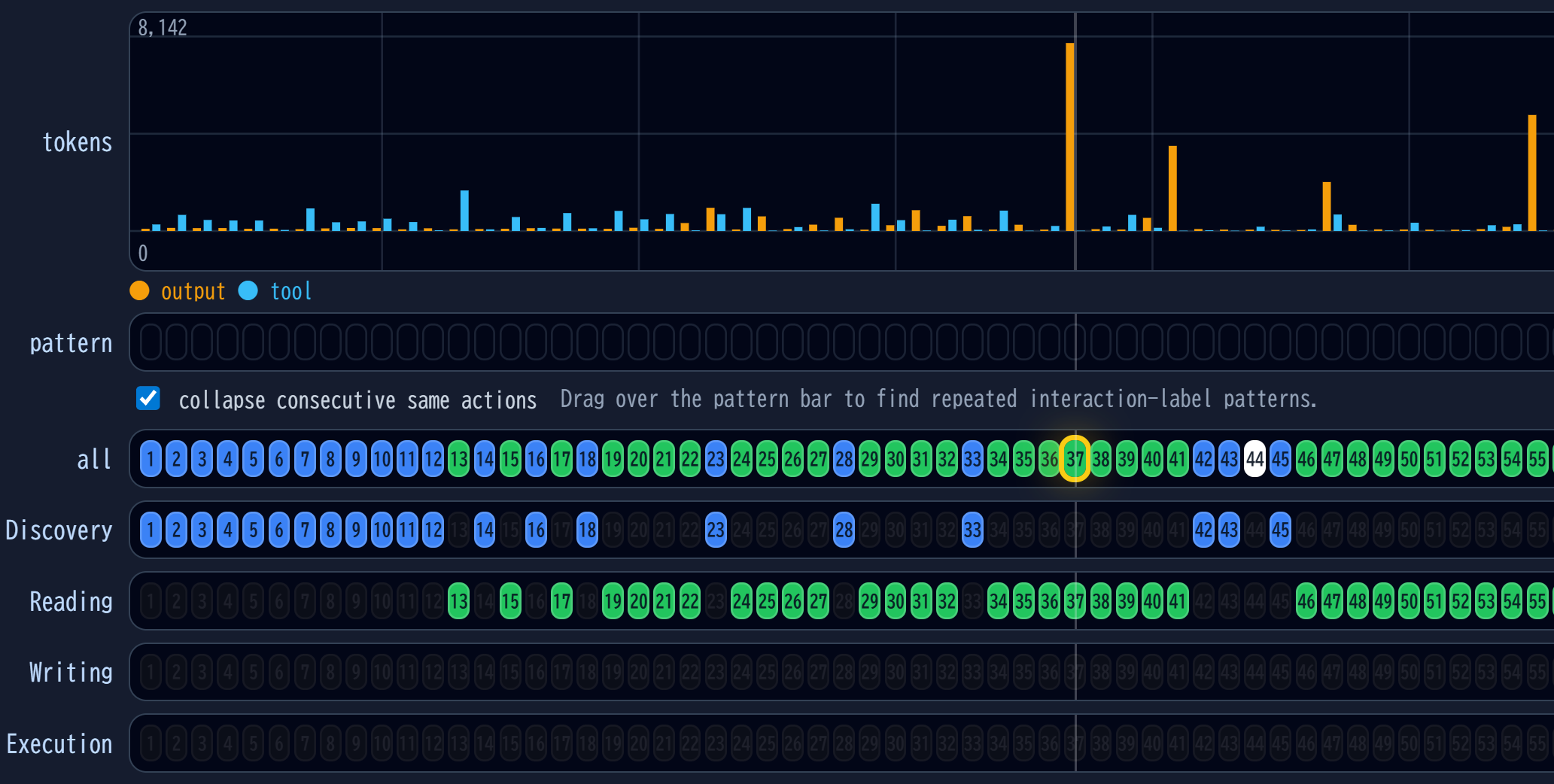}
  \caption{Case~2, Attempt~1. The largest spikes indicate that high token usage is mainly caused by long LLM output.}
  \label{fig:case2-timeline1}
\end{figure*}

\begin{figure*}[!t]
  \centering
  \includegraphics[
    width=\linewidth,
    trim=0pt 630pt 1500pt 0pt,
    clip
  ]{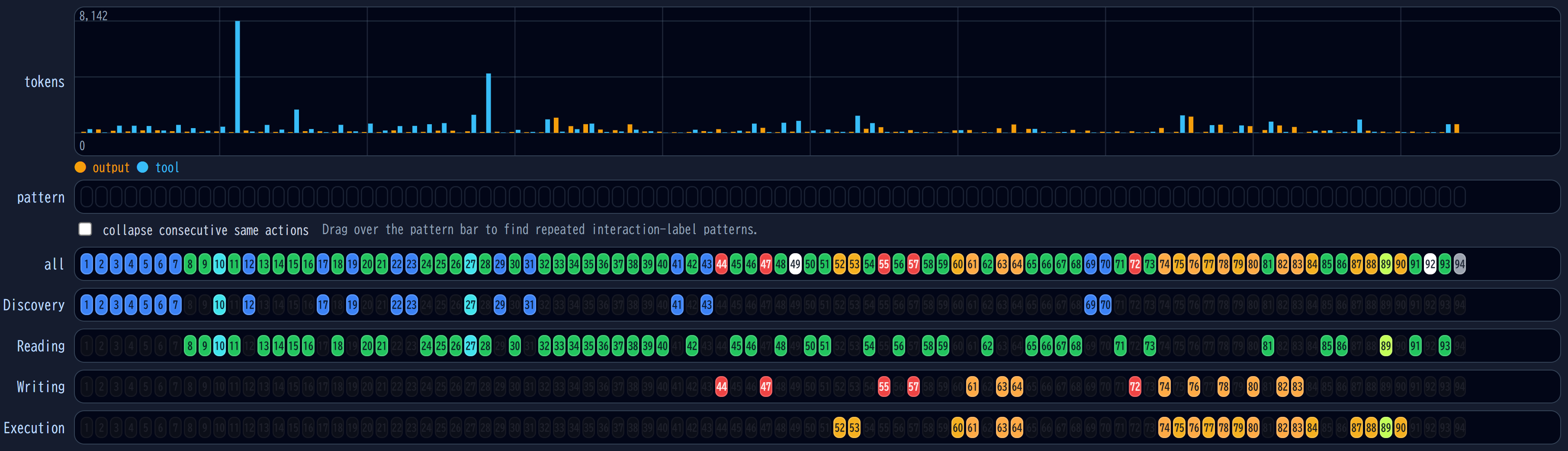}
  \caption{Case~2, Attempt~2. In contrast to Attempt~1, the largest spikes indicate that high token usage is mainly caused by large tool results.}
  \label{fig:case2-timeline2}
\end{figure*}

\subsection{User Interactivity}

The visualization allows users to move from the overview to detailed logs.
When a user hovers over an interaction, the visualization shows its representative label, such as \texttt{find | grep | head}, \texttt{git diff}, or \texttt{npm test}, and highlights other interactions with the same label.
This allows users to identify the concrete command or tool use behind each interaction and to notice repeated operations, such as recurring searches or repeated test executions, without placing all labels directly on the timeline.
Clicking an interaction opens a detail panel with the representative label, timing, token usage, tool results, and command outputs.
Thus, the visualization does not replace the original log; it helps users find the parts of the log that deserve closer inspection.

Users can also select a consecutive sequence of interactions and search for the same operation pattern elsewhere in the trajectory.
This supports visual inspection of recurring interaction patterns.


\section{Case Studies}
\label{sec:case-studies}

This section illustrates how Agent ATO compares selected runs under the same task, prompt, and environment.

\subsection{Study Design}
\label{subsec:case-study-design}

We do not assess whether the task was completed successfully. Instead, we examine process-level differences that are difficult to understand from the final patch alone. These include when the agent makes edits, whether it validates those edits, and which interactions use the most tokens.
In each case, we first use the timeline to identify a potential difference. We then inspect the detailed labels or logs to interpret that difference.

We used two real GitHub issues as target tasks: issue \texttt{\#28383}, ``Album ordering resets to stale order after opening asset viewer,'' in \textsc{immich}~\footnote{\url{https://github.com/immich-app/immich/issues/28383}}, and issue \texttt{\#74530}, ``\texttt{Formatting} section of Y-axis settings has scroll but it shouldn't,'' in \textsc{metabase}~\footnote{\url{https://github.com/metabase/metabase/issues/74530}}.
For each issue, we executed the same Pi coding agent 10 times with the same prompt and execution environment.
During each run, we collected the Pi event log, including turn boundaries, LLM messages, tool calls, bash commands, tool results, token usage, and timestamps.
Each log was then reconstructed as an interaction sequence using the method described in Section~\ref{2_approach}.
We have made the replication package for this experiment publicly available on GitHub.~\footnote{\url{https://anonymous.4open.science/r/VISSOFT2026ReplicationPackage-0E76}}

\subsection{Case 1: Comparing Trajectories for the Same Issue}
\label{subsec:case1}

The first case study targets the \textsc{immich} issue \texttt{\#28383}, a client-side state management and navigation bug.

Across the 10 executions, most attempts reached files related to the issue, but their repair strategies were not uniform.
We therefore focus on four attempts that treated the bug mainly as a page-state problem and modified overlapping locations.
This subset lets us compare process differences among executions that followed a similar repair direction.

Figs.~\ref{fig:case1-timeline1} and~\ref{fig:case1-timeline2} compare two representative attempts from this subset.
In both attempts, Writing interactions appear after earlier Discovery and Reading interactions, indicating that the agent gathered codebase context before editing.
The difference appears in the relationship between the Writing and Execution timelines.
In Fig.~\ref{fig:case1-timeline1}, red Writing interactions are repeatedly followed by yellow Execution interactions, especially in the latter half of the trajectory.
In Fig.~\ref{fig:case1-timeline2}, several Writing interactions appear, but the Execution timeline does not show corresponding validation after these edits.

Inspecting the representative labels of the Execution interactions in Fig.~\ref{fig:case1-timeline1} showed that these executions were test commands.
Thus, the visualization exposes a process difference that is not captured by the fact that both attempts pursued a similar repair direction: Attempt~1 formed edit-validation cycles, whereas Attempt~2 did not validate its edits within the observed trajectory.

Fig.~\ref{fig:case1-timeline1} also shows that the first validation required more intermediate interactions than later validations.
After the first Writing interaction, several additional interactions occurred before the first test execution; after later Writing interactions, test execution appeared immediately afterward.
This suggests that the agent first had to identify how to run a relevant test and then reused that command in later cycles.
This suggests a potential prompt-debugging approach: requiring early test identification and validation after editing.

\begin{tcolorbox}[title=Case 1 Finding]
Agent ATO exposed that, among attempts with similar repair directions, one repeatedly validated edits with test execution while another edited without subsequent validation.
\end{tcolorbox}

\subsection{Case 2: Token Consumption and Its Causes}
\label{subsec:case2}

The second case study targets the \textsc{metabase} issue \texttt{\#74530}, a UI bug in which the \texttt{Formatting} section of the Y-axis settings scrolls unnecessarily. This case focuses on the token-usage graph aligned with the interaction sequence. Among the 10 executions, we selected two attempts that both had high token usage but differed in source. Figs.~\ref{fig:case2-timeline1} and~\ref{fig:case2-timeline2} show the token-usage graphs for these attempts. Attempt~1 contains large spikes in LLM-output tokens, whereas Attempt~2 contains large spikes in tool-result tokens.

In Fig.~\ref{fig:case2-timeline1}, the dominant spikes are orange, indicating that the corresponding interactions consumed many LLM-output tokens. Inspecting the detailed log for the largest spike showed a long thinking segment in which the agent repeatedly reconsidered competing repair policies. This does not show that the final modification was wrong; rather, it identifies a token-efficiency issue. From a prompt-debugging perspective, such a pattern suggests instructing the agent to summarize candidate policies, select one, and continue concisely when repeated reconsideration occurs. At the system level, unusually long thinking segments could also trigger interruption, summarization, or user confirmation.

Fig.~\ref{fig:case2-timeline2} shows a different pattern: the largest spikes are blue, meaning that they come from tool-result tokens rather than LLM-output tokens. The detailed logs showed that these spikes were caused by large Reading results from file inspection. This is not necessarily an error, because the relevant information may reside in a large file. However, it suggests different interventions: prompts can ask the agent to check file size and read partial ranges, while agent designs can search for relevant identifiers first, read only necessary ranges, or summarize long tool results before continuing.

\begin{tcolorbox}[title=Case 2 Finding]
Agent ATO showed that high token usage can arise from different sources: long LLM-output segments suggest reasoning-control issues, whereas large tool results suggest context-management or partial-reading issues.
\end{tcolorbox}

\section{Conclusion and Future Work}
\label{sec:conclusion}

In this paper, we proposed Agent ATO, a timeline-based visualization for inspecting AI coding agent trajectories reconstructed from console logs. Agent ATO combines an all-interaction timeline, filtered timelines for Discovery, Reading, Writing, and Execution, and an aligned token-usage graph. Its goal is not to infer hidden intentions or automatically diagnose failures, but to help users observe what the agent did, in what order, and where operation-level or token-level differences occur.

Our case studies showed two uses of this representation. First, the timelines exposed differences in validation behavior, distinguishing repeated edit-validation cycles from edits without subsequent test execution. Second, the token-usage graph separated high token usage caused by long LLM output from high token usage caused by large tool results. These results suggest that trajectory visualization can guide detailed log inspection and support prompt-debugging and agent-design discussions.

Current evidence is limited to one agent, two tasks, selected runs, and unvalidated rule-based tags. Future work will evaluate tagging accuracy, compare Agent ATO with raw-log inspection, and study whether trajectory observations lead to effective prompt or agent revisions.

\section*{Acknowledgment}
This work was supported by JSPS KAKENHI Grant Number JP24H00692, JP25K03102, JP26K02889, JP26H02500, JP23K28065.

\bibliographystyle{IEEEtran}
\bibliography{bibliography}
\end{document}